\documentclass[10pt,letterpaper]{article}
\usepackage{opex3}
\usepackage{times,epsfig,epstopdf}
\usepackage{srcltx}
\usepackage{psfrag}
\usepackage{epsfig}
\usepackage{color}
\usepackage[tbtags,sumlimits,nointlimits,reqno]{amsmath}
\usepackage{amssymb}
\usepackage{color}
\usepackage{float}
\usepackage{subfigure}
\usepackage{multicol}
\usepackage{arydshln}
\usepackage{cite}
\usepackage[usenames,dvipsnames]{xcolor}

\renewcommand{\textcolor}[2]{#2}

\begin{document}

%%%%%%%%%%%%%%%%%% title page information %%%%%%%%%%%%%%%%%%
\title{Spontaneous Emission and Light Extraction Enhancement of Light Emitting Diode Using Partially-Reflecting Metasurface Cavity (PRMC)}

\author{Luzhou Chen$^1$, Karim Achouri$^1$, Themos Kallos$^2$ and Christophe~Caloz$^{1*}$}

\address{
$^1$\'{E}cole Polytechnique de Montr\'{e}al, Montr\'{e}al, Qu\'{e}bec H3T 1J4, Canada.
\\ $^2$ Metamaterial Technologies Inc., Dartmouth, Nova Scotia, B2Y 4M9, Canada.
\email{$^*$christophe.caloz@polymtl.ca}
}


\begin{abstract}
The enhancement of the power conversion efficiency~(PCE), and subsequent reduction of cost, of light emitting diodes~(LEDs) is of crucial importance in the current lightening market. For this reason, we propose here a PCE-enhanced LED architecture, based on a partially-reflecting metasurface cavity~(PRMC) structure. This structure simultaneously enhances the light extraction efficiency~(LEE) and the spontaneous emission rate~(SER) of the LED by enforcing the emitted light to radiate perpendicularly to the device, so as to suppress wave trapping and enhance field confinement near the emitter, while ensuring cavity resonance matching and maximal constructive field interference. The PRMC structure is designed using a recent surface susceptibility metasurface synthesis technique. A PRMC blue LED design is presented and demonstrated by full-wave simulation to provide LEE and SER enhancements by factors 4.0 and 1.9, respectively, which correspond to PCE enhancement factors of 6.2, 5.2 and 4.5 for IQEs of 0.25, 0.5 and 0.75, respectively, suggesting that the PRMC concept has a promising potential in LED technology.
\end{abstract}


%%%%%%%%%%%%%%%%%%%%%%% References %%%%%%%%%%%%%%%%%%%%%%%%%
%\bibliographystyle{osa}
%\bibliography{ReferenceList}

%%%%%%%%%%%%%%%%%%%%%%%%%%  body  %%%%%%%%%%%%%%%%%%%%%%%%%%
{\section{Introduction}}

Blue light emitting diodes (LEDs), which represent key constituents of commercial white LEDs, have spurred considerable interest in recent years due to their exceptionally low power consumption and long life span. However, boosting their power conversion efficiency (PCE) in order to reduce the cost remains a tantalizing goal. In LEDs, the PCE is proportional to the external quantum efficiency (EQE), which is the product of the internal quantum efficiency (IQE) and the light extraction efficiency (LEE)~\cite{EQE_IQE}. Since the IQE is proportional to the ratio of the number of radiated recombined electron-hole~(e-h) pairs over the number of all the recombined e-h pairs, it highly depends on the spontaneous emission rate (SER). Specifically, the enhancements of IQE and SER are related as~\cite{IQE_SER}

\begin{equation}
\label{eq:IQE_enhance}
\frac{IQE'}{IQE}=\frac{1}{1+\frac{SER}{SER'}(\frac{1}{IQE}-1)},
\end{equation}

\noindent where $IQE'$ and $IQE$ are the internal quantum efficiencies of the modified and reference LEDs, respectively, while $SER'$ and $SER$ are the spontaneous emission rates of the modified and reference LEDs, respectively. According to Eq.~\eqref{eq:IQE_enhance}, for a given value of $IQE$, higher $SER'/SER$ (or lower $SER/SER'$) results in higher $IQE'/IQE$, so that enhancing $IQE'/IQE$ essentially corresponds in enhancing $SER'/SER$. Thus, the PCE enhancement factor is

\begin{equation}
\label{eq:PCE_enhancement}
\frac{PCE'}{PCE}=\frac{LEE'}{LEE}\cdot\frac{IQE'}{IQE}=\frac{LEE'}{LEE}\cdot\frac{1}{1+\frac{SER}{SER'}(\frac{1}{IQE}-1)},
\end{equation}

\noindent and maximizing the PCE essentially requires maximizing both the LEE and the SER. This represents a challenging task because LEE and SER involve totally different physics, so that trade-offs may have to be made. LEE is typically impaired by the trapping of emitted photons within the active regions of the LED due to sharp refractive index contrast between the semiconductor medium (e.g. GaN, Si) and air. Many methods have been reported to mitigate this issue by increasing the escape cone using for instance distributed Bragg reflectors (DBR)~\cite{DBR_LED}, textured surfaces~\cite{textured_LED} or photonic crystals~\cite{PC_LED1,PC_LED2}. SER depends on the atomic structure of the emitter and on the density of electromagnetic modes of the environment~\cite{Novotny}. While the former is fixed for a given material, the latter may be enhanced by some reported methods such as coupling dipolar emission modes to resonant cavity modes~\cite{cavity_LED} or by matching the emission frequency to surface plasmon polaritons~\cite{SPP_LED}.

Metamaterials, which are artificial materials consisting of subwavelength arrangements of scattering inclusions in a host medium, can transform electromagnetic fields by changing their phase, amplitude and polarization in specified manners, and hence support many novel phenomena such as for instance negative refraction~\cite{negative_refraction}, cloaking~\cite{cloaking}, and light freezing~\cite{stop_light}. In particular, multilayered hyperbolic metamaterials have been recently reported as a novel approach for enhancing SER due to the infinite local density of state (LDOS) provided by their theoretically unlimited dispersion volume~\cite{HMM}. However, this approach is practically limited by the restricted thinness of the layers, causing Bragg scattering dispersion restriction, and requires complex designs, such as for instance grating structures~\cite{HMM_LED1,HMM_LED2}, to transform intrinsic lateral emission into vertical radiation for high LEE.

Metasurfaces are much easier to fabricate than their 3D metamaterial counterparts and may perform a great diversity of electromagnetic transformations, such as for instance generalized refraction~\cite{MS_generalized} and vortex wave generation~\cite{MS_vortex}. They are particularly suited to control fields in layered structures, since they can be conveniently stacked over such structures. In this work, we propose, synthesize and numerically demonstrate a novel blue LED structure with two metasurfaces forming a partially-reflecting metasurface cavity (PRMC) and leading to maximal PCE. This is achieved by simultaneously enhancing the LEE, by suppressing wave trapping, and the SER, by concentrating fields near the emitter.

\vspace{5mm}{\section{Principle of the PRMC LED}}

\begin{figure}
\centering\includegraphics[width=11cm]{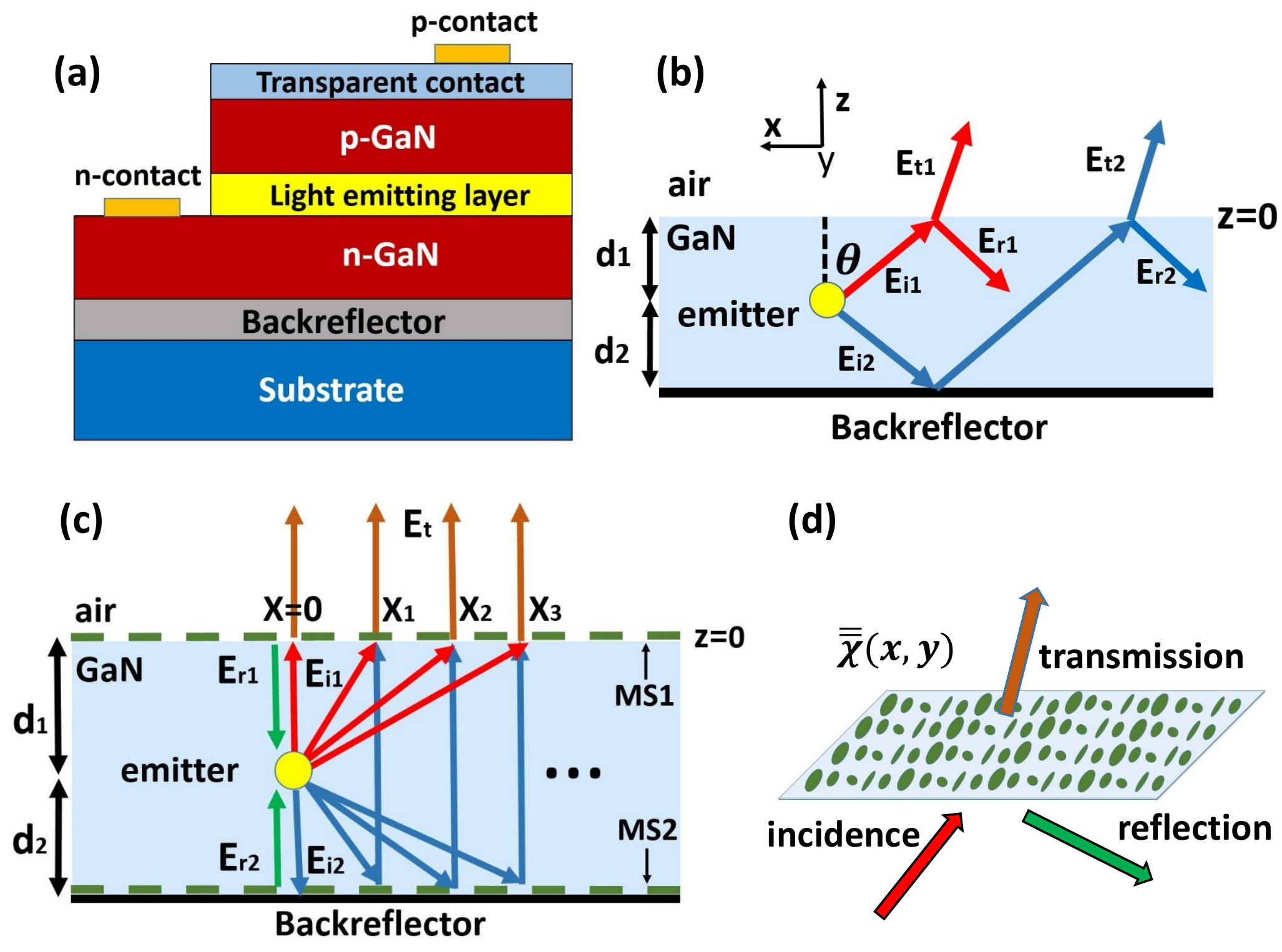}
\caption{\label{fig:illustration}(a)~Multilayer configuration (not to scale) of a typical GaN LED, with electrodes, transparent contact layer, light emitting layer sandwiched between the p-GaN and n-GaN layers, backreflector and substrate. (b)~Simplified LED of (a) with unique GaN layer, dipole emitter, and perfect backreflector. (c)~Proposed partially-reflecting metasurface cavity (PRMC) LED structure with front metasurface (MS1) and back metasurface (MS2), GaN layer, dipole emitter and perfect backreflector. (d)~Modeling of each metasurface by a sheet of tensorial surface susceptiblity, $\bar{\bar{\chi}}(x,y)$, synthesized for arbitrary incident, reflected and transmitted fields~\cite{MS_susceptibility}.}
\end{figure}

We consider here the case of Gallium Nitride (GaN) LEDs, given their prominent role in commercial applications, but the proposed concept naturally applies to the other semiconductors, such as Silicon (Si), with parameter adjustments. Figure~\ref{fig:illustration}(a) shows the cross-section of a typical GaN LED, with pad contacts, transparent contact layer, light emitting layer sandwiched between the p-GaN and n-GaN layers, metal backreflector and substrate. Since the transparent contact layer usually has a refractive index close to that of air, it may be ignored. The light emitting layer thickness, being in the order of a few tens of nanometers, may also be ignored. The backreflector, which is highly reflective (typically DBR or silver layer), may be modeled as a perfect electric conductor (PEC) sheet. Moreover, we shall replace the light emitting layer by a simple dipole to simplify the forthcoming synthesis of the metasurfaces. Finally, since the refractive indices of p-GaN and n-GaN are very close to that of pure GaN, they can be merged in a single one with the index of pure GaN. As a result of all these simplifications, the structure in Fig.~\ref{fig:illustration}(a), for simulation purpose, reduces to that shown in Fig.~\ref{fig:illustration}(b), consisting to a unique layer of pure GaN with a dipole quantum emitter position at distances $d_1$ and $d_2$ from the air interface and backreflector, respectively, representing an infinitesimal element of the light emitting layer, on a perfect backreflector.

Assuming blue LED ($\lambda_0 = 490$~$\rm nm$), $\lambda_\text{GaN} = 490/\sqrt{\varepsilon_\text{GaN}} \approx 200$ $\rm nm$ ($\varepsilon_\text{GaN} = 6.004$) is much smaller than the total thickness of the GaN layer ($\sim 1-10~\mu$m) and therefore ray optics may be applied as a heuristic tool to describe electromagnetic field propagation in the structure. As depicted in Fig.~\ref{fig:illustration}(b), the field emitted by the dipole may be decomposed, for later comparison, into a direct field, $E_{i1}$, directly propagating towards the GaN-air interface and scattered at it, and an indirect field, $E_{i2}$, first reflected on the backreflector. Note that $E_{i2}$ can be considered as the field produced by a mirror dipole according to image theory. Due to the large contrast of refractive indices, the escape angle from GaN to air, according to Snell law, is only $\theta_\text{c}=\sin^{-1}(1/n_\text{GaN})=24.1^\circ$, which leads to massive guided-mode loss, with most of the dipole emission energy trapped within the GaN layer and little radiated out. As a result, such a configuration suffers from poor LEE, and hence poor PCE. The small amount of radiated energy may be maximized by setting the parameters $(d_1,~d_2)$ so as to satisfy the Fabry-Perot conditions for both the direct and indirect waves, but this is largely insufficient given the amount of lost energy in trapped waves.

The proposed PRMC LED is shown in Fig.~\ref{fig:illustration}(c) with metasurface model represented in Fig.~\ref{fig:illustration}(d). The front (top in the figure) metasurface (MS1) is a partially-reflecting metasurface placed at the GaN-air interface, while the back (bottom in the figure) metasurface (MS2) is a completely reflective metasurface placed between the GaN layer and the backreflector. To avoid plasmonic losses, the metasurfaces should be ideally implemented in all-dielectric metasurface technology~\cite{MM_dielectric}. The PRMC structure essentially operates as follows. MS1 normally refracts the direct dipole field, $E_{i1}$, for all incidence angles, into part of the plane-wave transmitted field, $E_t$, and normally reflects $E_{i1}$ into the field $E_{r1}$. MS2 normally reflects the indirect field, $E_{i2}$, for all incidence angles, into the field $E_{r2}$, and transmits this field to complete $E_t$. Since all the light rays, including those above $\theta_\text{c}$ in the case of Fig.~\ref{fig:illustration}(b), are converted to normal rays, wave trapping within the GaN layer is completely suppressed, so that a much higher LEE level may be expected. On the other hand, SER maximization requires the dipole emitter to be placed at an antinode of the standing wave formed in the PRMC along the $z$-direction to leverage the Purcell effect~\cite{Purcell}. If the reflection phase shifts induced by MS1 ($\phi_{r1}$) and MS2 ($\phi_{r2}$) were zero, this would correspond to simultaneously satisfying the resonance condition

\begin{subequations}
\label{eq:pos_SER}
\begin{equation}
\label{eq:pos_d1d2}
d_1+d_2=m\lambda_\text{GaN}/2\quad (m~\text{integer}),
\end{equation}
and the antinode position condition, which taking into account~\eqref{eq:pos_d1d2}, reads
\begin{equation}
d_1=(n/2+1/4)\lambda_\text{GaN}\quad(n~\text{integer}).
\end{equation}
\end{subequations}

\noindent However, the situation is more complicated because one also needs to maximize the constructive interference of the waves transmitted by the direct and indirect fields, in $E_t$, for maximal LEE. In terms of the ray scheme, this would imply the constructive interference of an infinite number of direct (d) and indirect (i) wave sets, such as for instance, with reference to Fig.~\ref{fig:illustration}(c),
%
%\begin{align}
\begin{subequations}
\label{eq:ci_LEE}
\begin{equation}
\sqrt{d_2^2+x_m^2}+d_1+d_2-\sqrt{d_1^2+x_m^2}=q_1\lambda_\text{GaN}\quad\text{(i-d)},
\end{equation}
\begin{equation}
\sqrt{d_1^2+x_m^2}-\sqrt{d_1^2+x_n^2}=q_2\lambda_\text{GaN}\quad\text{(d-d)},
\end{equation}
\begin{equation}
\sqrt{d_2^2+x_m^2}-\sqrt{d_2^2+x_n^2}=q_3\lambda_\text{GaN}\quad\text{(i-i)},
\end{equation}
\begin{equation*}
\text{etc.},
\end{equation*}
\end{subequations}
%\end{align}
%
\noindent where $m,n$ and $q_1,q_2,q_3,\ldots$ are integers. Naturally, these conditions cannot be all simultaneously satisfied, given the finite number of design parameters, and they are electromagnetically not exact anyways. However, they suggest that optimal LEE should correspond to a design where fields would form light rings at the positions of maximal constructive interference (points $x_m,x_n,\ldots$). In a real design, this optimum will be found by full-wave electromagnetic simulation. Specifically, we will enforce $\phi_{r1}=0$ and tune $\phi_{r2}$ and $\phi_{t}$ (transmission phase of MS1) so as to achieve an optimal design in terms of both LEE and SER, i.e. to maximize the overall LED PCE. The full-wave simulations will be performed using the COMSOL finite element method (FEM) commercial software, where the metasurface is modeled as a deeply sub-wavelength slab with volume susceptibility $\bar{\bar{\chi}}_\text{vol}=\bar{\bar{\chi}}/\delta$, where $\delta$ is the thickness of the slab~\cite{Yousef}. Given the high density of the FEM mesh required in the metasurface-modeling slab and around the (point) dipole, the actual 3D problem is not tractable on a standard computer. Therefore, it will be reduced to a 2D model ($\partial/\partial y=0$), without altering the key conclusions in terms of fundamental comparison between the bare LED and PRMC LED structures.

\vspace{5mm}{\section{Synthesis of the metasurfaces}}

In order to transform the incident field into reflected and transmitted fields as specified above, an efficient metasurface synthesis technique is required. Various metasurface synthesis methods have been recently reported, including methods based on polarizabilities~\cite{MS_polarizability}, impedance tensors~\cite{MS_impedance}, momentum transformation~\cite{MS_momentum} and surface susceptibility tensor~\cite{MS_susceptibility}. We will use here the last of these techniques, which can handle full vectorial electromagnetic fields and provide closed-form solutions for quasi-arbitrary wave transformations. This technique will be applied to determine the initial surface susceptibility functions of the metasurfaces MS1 and MS2 in Fig.~\ref{fig:illustration}(d), namely $\bar{\bar{\chi}}_1(x)$ and $\bar{\bar{\chi}}_2(x)$, for generating the normal scattered waves indicated Fig.~\ref{fig:illustration}(c). Being deeply sub-wavelength in thickness, a metasurface can be considered as a zero-thickness discontinuity of space, and may then be generally modeled by the Generalized Sheet Transition Conditions (GSTCs)~\cite{Idemen,MS_susceptibility}
\begin{subequations}
\label{eq:GSTC}
\begin{equation}
\hat{z}\times\Delta \textbf{\emph{H}} = j\omega \textbf{\emph{P}}_{\parallel}-\hat{z}\times \nabla_{\parallel}M_z,
\end{equation}
\begin{equation}
\Delta \textbf{\emph{E}}\times \hat{z} = j\omega\mu_0 \textbf{\emph{M}}_{\parallel}-\nabla_{\parallel}(\frac{P_z}{\varepsilon_0})\times \hat{z},
\end{equation}
\end{subequations}
\noindent where the symbol $\Delta$ refers to the difference of the electromagnetic fields $\emph{\textbf{E, H}}$ between the two sides of the metasurface (e.g. $\Delta E^u = E^u_t-(E^u_i + E^u_r),~u=x,y,z$), and $\emph{\textbf{P}}$ and $\emph{\textbf{M}}$ are the electric and magnetic polarization densities, respectively.

In general, a metasurface may induce electromagnetic couplings, in which case it is described by the bianisotropic constitutive relations $\emph{\textbf{P}}=\varepsilon_0\bar{\bar{\chi}}_{ee}\emph{\textbf{E}}_\text{av}+\sqrt{\mu_0\varepsilon_0}\bar{\bar{\chi}}_{em}\emph{\textbf{H}}_\text{av}$ and $\emph{\textbf{M}}=\bar{\bar{\chi}}_{mm}\emph{\textbf{H}}_\text{av}+\sqrt{\varepsilon_0 / \mu_0}\bar{\bar{\chi}}_{me}\emph{\textbf{E}}_\text{av}$, where the subscript ``\text{av}'' denotes the average of the fields at the two sides of the metasurfaces (e.g. $E^u_\text{av} = [E^u_t +(E^u_i + E^u_r)]/2,~u=x,y,z$). However, since active layer materials (e.g. GaN, Si) in optoelectronic devices are typically isotropic, we set $\bar{\bar{\chi}}_{em}=\bar{\bar{\chi}}_{me}=0$ and $\chi^{xy}_{ee}=\chi^{yx}_{ee}=\chi^{xy}_{mm}=\chi^{yx}_{mm}=0$. Furthermore, we will seek the simplest possible design, assuming $P_z=M_z=0$, in which case Eqs.~\eqref{eq:GSTC} yield the closed-form surface susceptibilities
\begin{subequations}
\label{eq:susceptibilities}
\begin{equation}
\chi^{xx}_{ee} =\frac{-\Delta H^y}{j\omega\varepsilon_0 E^x_\text{av}},\quad
\chi^{yy}_{ee} =\frac{\Delta H^x}{j\omega\varepsilon_0 E^y_\text{av}},
\end{equation}

\begin{equation}
\chi^{xx}_{mm} =\frac{\Delta E^y}{j\omega\mu_0 H^x_\text{av}},\quad
\chi^{yy}_{mm} =\frac{-\Delta E^x}{j\omega\mu_0 H^y_\text{av}}.
\end{equation}
\end{subequations}

In the 2D model represented by Fig.~\ref{fig:illustration}, we consider a $y$-directed electric source (infinite line source), corresponding to s-polarization. This is in fact still a complicated problem, where complex multiple scattering coupling exists between the two metasurfaces. Therefore, as an initial design approach, we shall consider the two surfaces as uncoupled and only transforming all incident fields into normal fields. Coupling coherence will be ensured by setting the parameters $d_1$ and $d_2$, as previously mentioned, and by full-wave optimizing the entire structure. The corresponding surface susceptibilities in Eqs.~\eqref{eq:susceptibilities} for MS1 and MS2 take then the explicit forms
\begin{subequations}
\label{eq:chi1}
\begin{equation}
\chi^{yy}_{ee1}=\frac{2}{j\omega\varepsilon_0}\frac{H^x_t-(H^x_{i1}+H^x_{r1})}{E^y_t+(E^y_{i1}+E^y_{r1})},\label{equ:chi_yy_ee1}
\end{equation}
\begin{equation}
\chi^{xx}_{mm1}=\frac{2}{j\omega\mu_0}\frac{E^y_t-(E^y_{i1}+E^y_{r1})}{H^x_t+(H^x_{i1}+H^x_{r1})},\label{equ:chi_xx_mm1}
\end{equation}
\end{subequations}
and
\begin{subequations}
\label{eq:chi2}
\begin{equation}
\chi^{yy}_{ee2}=\frac{2}{j\omega\varepsilon_0}\frac{H^x_{i2}+H^x_{r2}}{E^y_{i2}+E^y_{r2}},\label{equ:chi_yy_ee2}
\end{equation}
\begin{equation}
\chi^{xx}_{mm2}=\frac{2}{j\omega\mu_0}\frac{E^y_{i2}+E^y_{r2}}{H^x_{i2}+H^x_{r2}},\label{equ:chi_xx_mm2}
\end{equation}
\end{subequations}
\noindent respectively, with $\chi^{xx}_{ee1}=\chi^{yy}_{mm1}=\chi^{xx}_{ee2}=\chi^{yy}_{mm2}=0$. Similar relations are naturally obtained for the case of a $y$-directed magnetic source, corresponding to p-polarization. Note that, given the complex nature of the involved fields, the surface susceptibilities in Eqs.~\eqref{eq:chi1} and Eqs.~\eqref{eq:chi2} will strongly depend on $x$ ($\chi=\chi(x)$), so that the response of the PRMC structure will strongly depend on the relative position of the metasurface with respect to the emitter. In order to find the scattering particles producing the synthesized susceptibilities in Eqs.~\eqref{eq:chi1} and Eqs.~\eqref{eq:chi2}, one first converts surface susceptibilities into scattering parameters~(S-parameters)~\cite{MS_susceptibility}, which yields in our problem
\begin{subequations}
\label{eq:S_parameters}

\begin{equation}
S_{\text{GaN}\rightarrow\text{air}}^{(k)}=\frac{-2\eta_0\sqrt{\eta_{\text{GaN}}/\eta_0}(4+\chi^{xx}_{ee(k)}\chi^{yy}_{mm(k)}\varepsilon_0\mu_0\omega^2)}
{-4(\eta_0+\eta_\text{GaN})-4i(\chi^{xx}_{ee(k)}\varepsilon_0\eta_0\eta_\text{GaN}+\chi^{yy}_{mm(k)}\mu_0)\omega+\chi^{xx}_{ee(k)}\chi^{yy}_{mm(k)}\varepsilon_0(\eta_0+\eta_\text{GaN})\mu_0\omega^2},\\
\end{equation}

\begin{equation}
S_{\text{GaN}\rightarrow\text{GaN}}^{(k)}=\frac{4(\eta_\text{GaN}-\eta_0)+4j(\chi^{xx}_{ee(k)}\varepsilon_0\eta_0\eta_\text{GaN}-\chi^{yy}_{mm(k)}\mu_0)\omega+\chi^{xx}_{ee(k)}\chi^{yy}_{mm(k)}\varepsilon_0(-\eta_\text{GaN}+\eta_0)\mu_0\omega^2}
{-4(\eta_0+\eta_\text{GaN})-4j(\chi^{xx}_{ee(k)}\varepsilon_0\eta_0\eta_\text{GaN}+\chi^{yy}_{mm(k)}\mu_0)\omega+\chi^{xx}_{ee(k)}\chi^{yy}_{mm(k)}\varepsilon_0(\eta_0+\eta_\text{GaN})\mu_0\omega^2},\\
\end{equation}

\end{subequations}
\noindent where $\eta_0$ and $\eta_\text{GaN}$ are the intrinsic impedances of air and GaN, respectively. The bracketed superscript denote the metasurface number, e.g. $S_{\text{GaN}\rightarrow\text{air}}^{(1)}$ is the transmission coefficient of MS1 from GaN to air and $S_{\text{GaN}\rightarrow\text{GaN}}^{(1)}$ is the reflection coefficient of MS1 at the GaN side. The S-parameters functions provide precious insight into the spatial variations of the surface susceptibilities, and hence on the realizability of the metasurface, which is ultimately discretized into sub-wavelength cells ($<\lambda_\text{eff}/5$), and allow one to determine the geometry of the scattering particles using standard parametric mapping~\cite{MS_susceptibility,MS_S_parameter}.

\vspace{5mm}{\section{LEE and SER computation}}

The LEE is simply calculated by integrating the radiated power along an $x$-directed line cut at one free-space wavelength above the structure. The SER of the dipole quantum emitter of the transition frequency~$\omega$ at point $\textbf{r}_0$ may be computed in terms of the dyadic Green function $\bar{\bar{\textbf{G}}}$ as~\cite{Novotny,SER_FDFD,SER_BEM}
\begin{equation}
\gamma(\textbf{r}_0,\omega)=\frac{2\omega^2}{\hbar\varepsilon_0 c^2}\langle \textbf{p}\cdot \rm Im[\bar{\bar{\textbf{G}}}(\textbf{r}_0,\textbf{r}_0;\omega)]\cdot\textbf{p}\rangle,
\end{equation}
\noindent where $\textbf{p}$ is the dipole moment of the quantum emitter, $c$ is the speed of light and $\hbar$ is the reduced Planck constant. In the considered 2D problem with $y$-polarized source [$\textbf{p}=p\delta(x)\delta(z+d_1)\hat{y}$], this expression reduces to
\begin{equation}
\label{eq:SER_comp}
\gamma(\textbf{r}_0,\omega)=\frac{2p^2\omega^2}{\hbar\varepsilon_0 c^2}{\rm Im [G_{\emph{yy}}(\textbf{r}_0,\textbf{r}_0;\omega)]},
\end{equation}

\noindent where the Green function in Eq.~\eqref{eq:SER_comp} is, by definition, simply the $y$-component of the electric field response to a point source oriented along the $y$-direction in the complete PRMC structure. This Green function is clearly not available in analytical form in such a complex problem, and it will therefore be computed numerically as the electric field $E_y$ produced by the point source $\textbf{p}=p\delta(x)\delta(z+d_1)\hat{y}$ in the forthcoming full-wave simulations.

\begin{figure}
\centering\includegraphics[width=9cm]{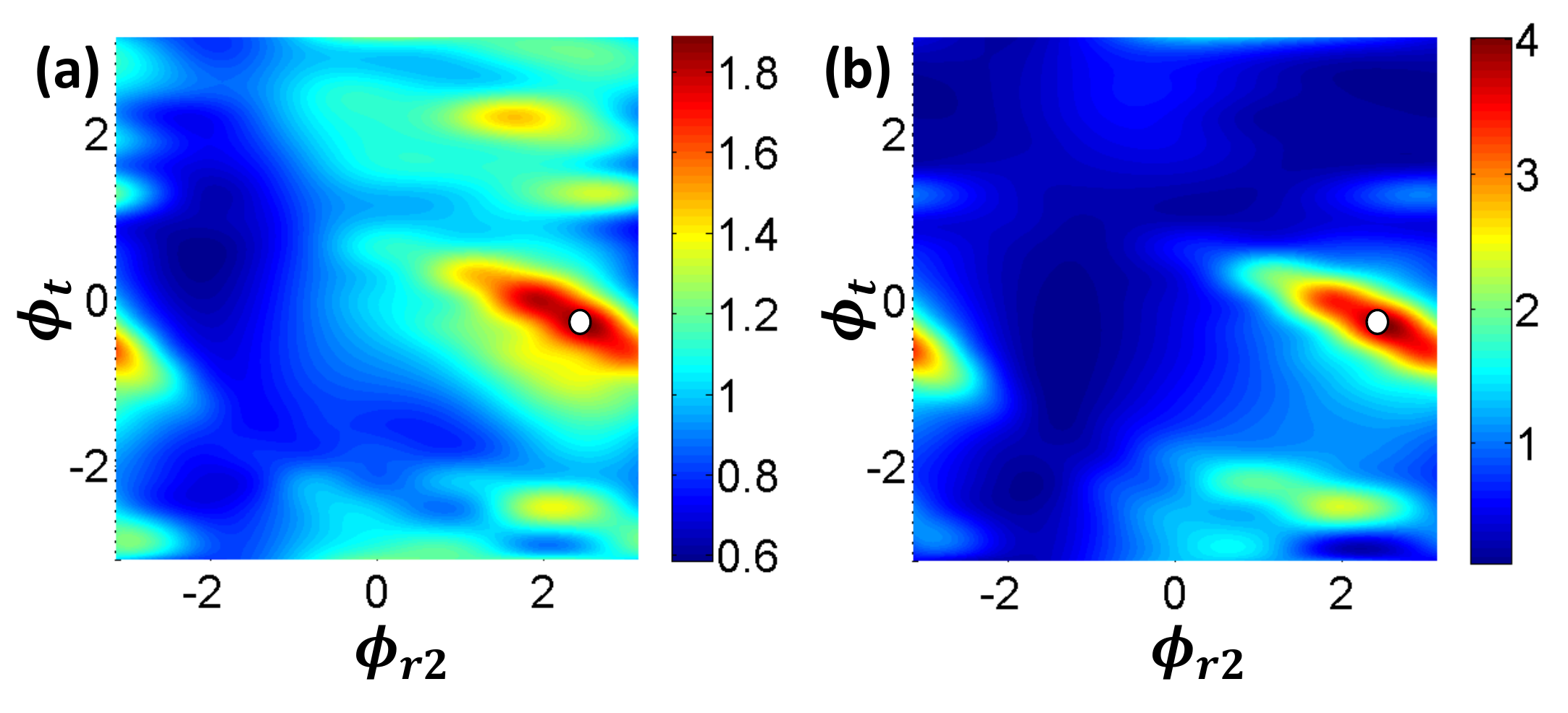}% Here is how to import EPS art
\caption{\label{fig:phase}Parametric optimization of the PRMC LED with respect to a bare LED in terms of the reflection phase of MS2, $\phi_{r2}$, and the transmission phase of MS1, $\phi_t$. (a)~SER enhancement factor. (b)~LEE enhancement factor. The white points indicate the phase pair providing simultaneous maximum SER and LEE enhancements, namely $(\phi_{r2},\phi_t)=(0.8\pi,-0.1\pi)$, as obtained after  optimizing the initial parameters in the design procedure described above.}
\end{figure}

\vspace{5mm}{\section{Numerical results and discussion}}

At this point, we may start the design of the PRMC LED structure. The overall procedure is as follows. First, the dipole position is set so as to simultaneously maximize the SER, by enforcing Eq.~\eqref{eq:pos_SER}, and the LEE, by enforcing Eq.~\eqref{eq:ci_LEE}, in the bare LED structure. This leads, for instance, to $(d_1,d_2)=(450,~450)$~nm. Second, the two metasurfaces are introduced in order to scatter all the rays perpendicularly to the LED structure, as shown in Fig.~\ref{fig:illustration}(c), to suppress wave trapping, so as to enhance the LEE, and also the SER as a result of higher field confinement near the emitter. As previously explained, the susceptibilities are first obtained as surface (zero-thickness) susceptibility functions, computed by Eq.~\eqref{eq:chi1} and Eq.~\eqref{eq:chi2}, $\chi(x)$, and next modeled in COMSOL as a deeply subwavelength slab (here 5~nm) with volume susceptibility $\chi_\text{vol}(x)$. This provides an initial design for the surface susceptibility functions of $x$ in~\eqref{eq:chi1} and~\eqref{eq:chi2}. Third, the reflection phase of MS1, $\phi_{r1}$, is set to zero [in $E_{r1}=E_{r10}(x)\exp(j\phi_{r1})$, $E_{r10}(x)=|E_{i1}|R$, $R$: local reflection coefficient], while the reflection phase of MS2, $\phi_{r2}$ [in $E_{r2}=E_{r20}(x)\exp(j\phi_{r2})$, $E_{r20}(x)=|E_{i2}|$], and the transmission phase of MS1, $\phi_{t}$ [in $E_{t}=E_{t0}(x)\exp(j\phi_{t})$, $E_{t0}(x)=|E_{i1}|T$], $T$: local reflection coefficient], are scanned for optimization, as shown in Fig.~\ref{fig:phase}. Applying Eq.~\eqref{eq:chi1} and Eq.~\eqref{eq:chi2} for the scanned values of $\phi_{r2}$ and $\phi_{t}$ eventually leads to the optimal pair ($\phi_{r2},\phi_{t}$) in terms of simultaneous LEE and SER maximization, which turns out, after adjusting $(d_1,d_2)$ to $(650,~450)$~nm (extra $\lambda_\text{GaN}$ added to $d_1$), to be equal to $(\phi_{r2},\phi_t)=(0.8\pi,-0.1\pi)$. This leads to adjusted surface susceptibilities corresponding to the final design. In case the synthesized susceptibilities exhibits a positive imaginary part [$\rm Im(\chi(x))>0$] in some areas, which would correspond to unpractical gain (active) metasurfaces, then one needs to sacrifice some SER or LEE to ensure a purely passive (and slightly lossy) design.

\begin{figure}
\centering\includegraphics[width=9cm]{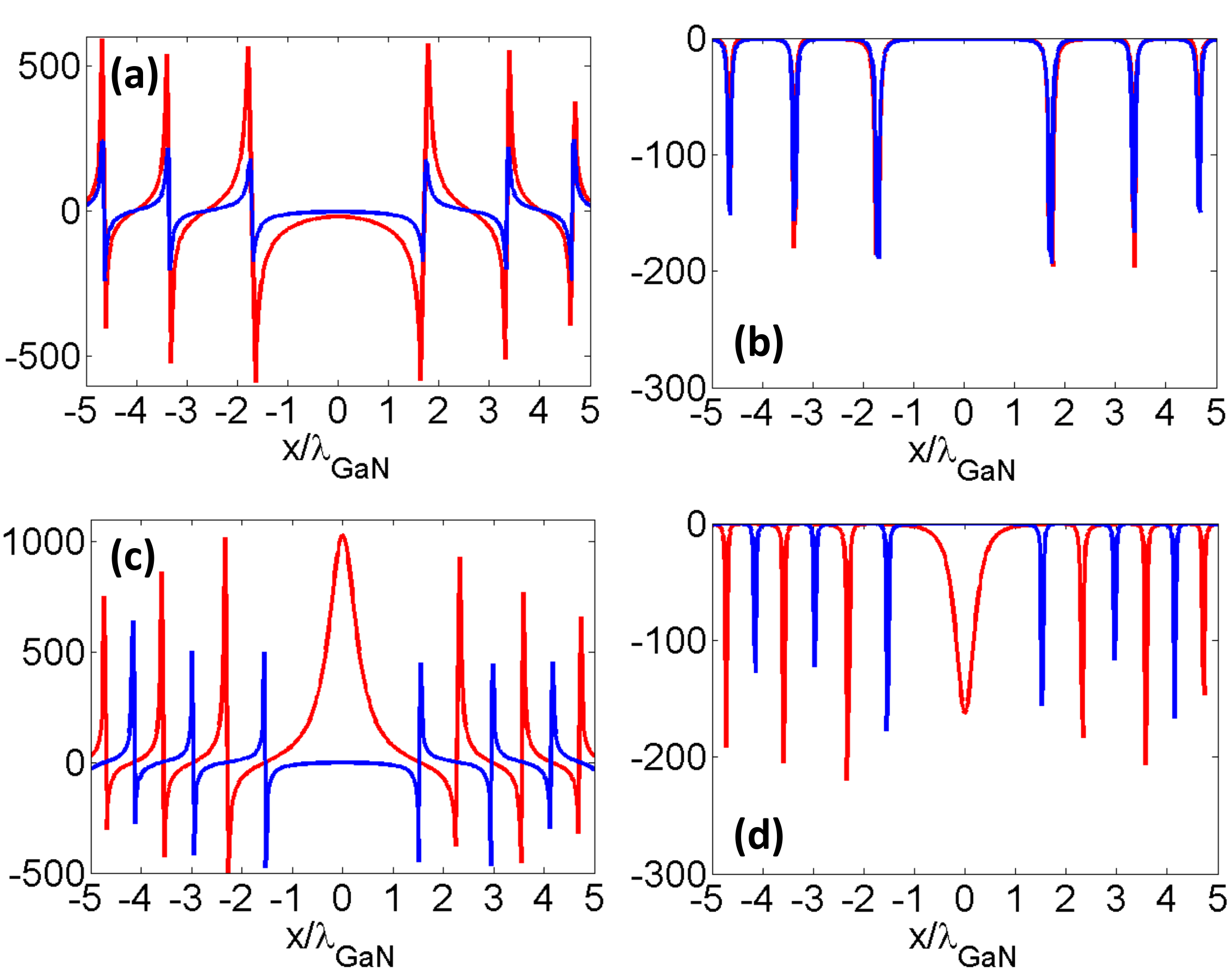}% Here is how to import EPS art
\caption{\label{fig:susceptibility}Synthesized electric (red) and magnetic (blue) surface susceptibilities of the two metasurfaces: (a)~real parts and (b)~imaginary parts for MS1; (c)~real parts and (d)~imaginary parts for MS2.}
\end{figure}

\begin{figure}
\centering\includegraphics[width=10cm]{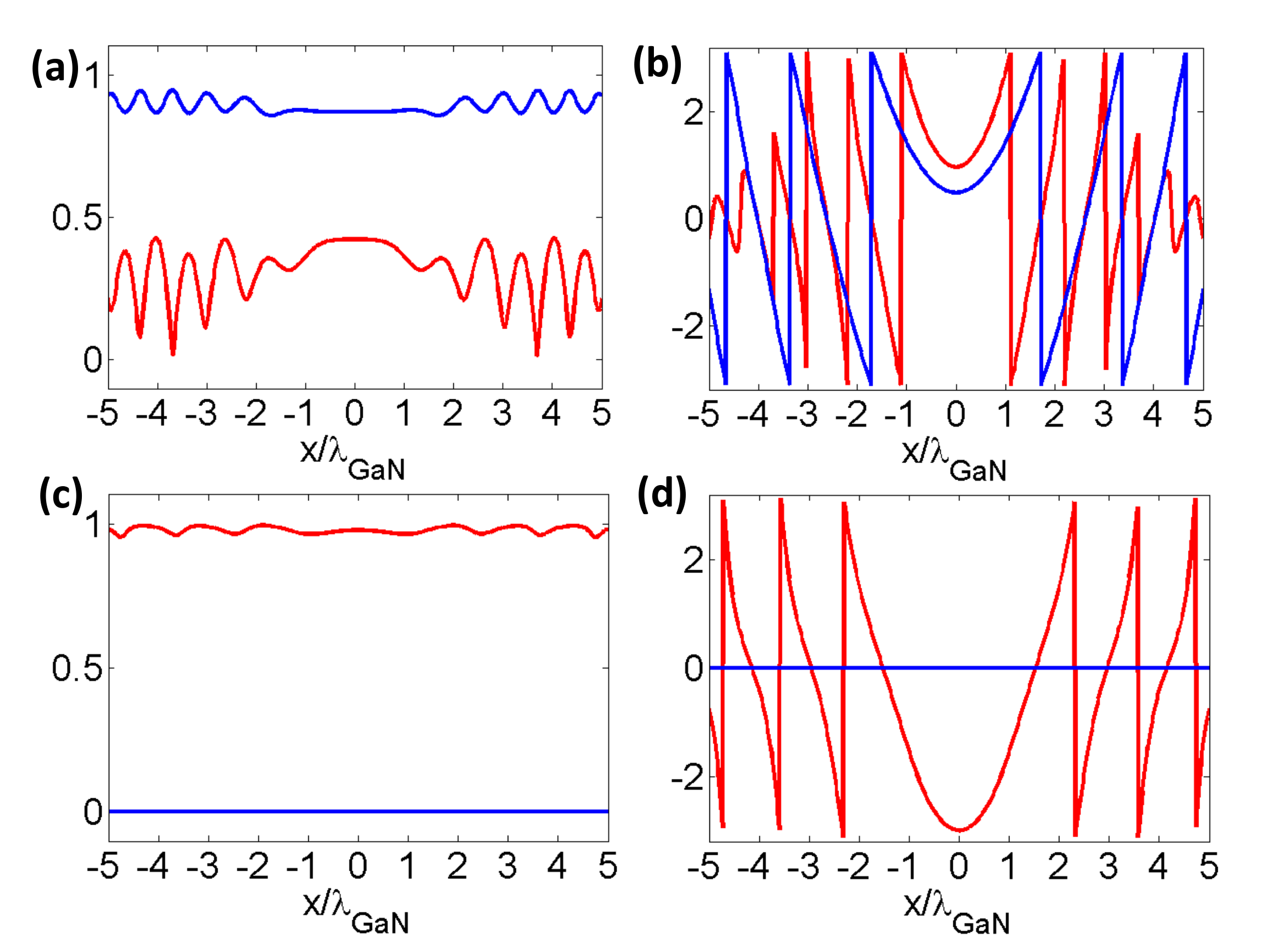}% Here is how to import EPS art
\caption{\label{fig:S-parameter}Scattering parameters obtained from Fig.~\ref{fig:susceptibility} with Eq.~\eqref{eq:S_parameters}: (a)~absolute values and (b)~phases of the reflection~(red) and transmission~(blue) coefficients of MS1; (c)~absolute values and (d)~phases of the reflection~(red) and transmission~(blue) coefficients of MS2.}
\end{figure}

Figure~\ref{fig:susceptibility} plots the synthesized surface susceptibility functions of the two metasurfaces versus position $x$. The origin corresponds to the position of the emitter. Note that the imaginary parts of all the susceptibility functions are always negative, as required for a purely passive design. The corresponding scattering parameters, obtained by applying Eq.~\eqref{eq:S_parameters} to the results in Fig.~\ref{fig:susceptibility}, are plotted in Fig.~\ref{fig:S-parameter}. It may be seen that the spatial variations of the scattering parameters are relatively small on the scale of the GaN wavelength, indicating that corresponding susceptibility responses should be easily achievable with practical scattering particles, whose size is typically in the order of $\lambda_\text{eff}/5$, using conventional mapping techniques~\cite{MS_susceptibility,MS_S_parameter}.

\begin{figure}
\centering\includegraphics[width=12cm]{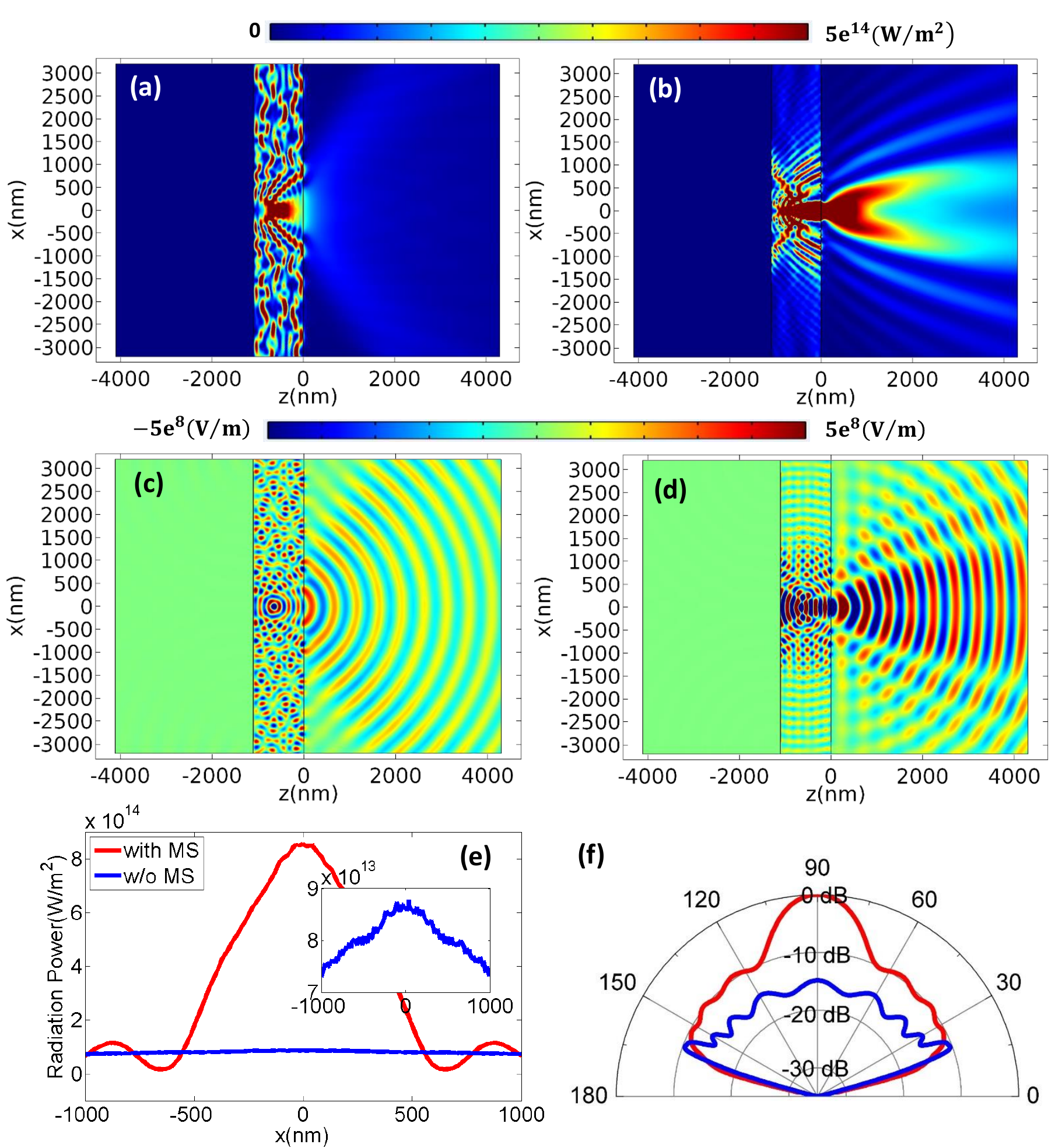}% Here is how to import EPS art
\caption{\label{fig:power}Comparison (full-wave FEM simulation) of the optical performance of the bare LED [Fig.~\ref{fig:illustration}(b)] and PRMC LED [Fig.~\ref{fig:illustration}(c)] structures. (a)~Power distribution for the bare LED in the longitudinal ($xz$-plane). (b)~Idem for the PRMC LED. (c)~Electric field ($E_y$) distribution for the bare LED in the same plane. (d)~Idem for the PRMC LED. In (a) to (d), the simulation region is surrounded by perfect matched layers (PMLs) while the perfect backreflector is modelled by a perfect electric conductor (PEC) sheet. In (a) and (c), from left to right are the substrate (isolated by PEC), PEC backreflector, GaN layer and air, while in (b) and (d), from left to right are the substrate (isolated by PEC), PEC backreflector, back metasurface~(MS2), GaN layer, front metasurface~(MS1) and air. The PEC sheet and metasurfaces are not visible due to their zero and extremely small (5nm) thicknesses, respectively. (e)~Light extraction power of the bare LED (blue line) and PRMC LED (red line) along a $x$-directed cut line at one free-space wavelength above the LED surface [extracted from (a) and (b)]. The inset shows a zoomed view of the bare LED result. (f)~Far-field radiation patterns of the bare LED (blue line) and PRMC LED (red line) in dB scale. Both lines are normalized to the maximum value of PRMC LED result. }
\end{figure}

Figure~\ref{fig:power} compares the optical performance of the bare LED and PRMC LED structures. Figures~\ref{fig:power}(a) and~(b) show the power distribution in the bare LED and PRMC LED structures, respectively, while Figs.~\ref{fig:power}(c) and~(d) show the electric field distribution in the bare LED and PRMC LED structures, respectively. As expected, in the case of the bare LED, the abrupt change of refractive index from GaN to air leads to a very small escape cone, so that most of the dipole emitted energy is trapped as guided modes propagating along the GaN layer, as seen in Figs.~\ref{fig:power}(a) and~(c). This issue is essentially solved in the PRMC, as seen in Figs.~\ref{fig:power}(b) and~(d), where the metasurfaces scatter all the emitted waves perpendicularly to the LED, leading to both enhanced LEE, via increased power extraction, and enhanced SER, via increased field confinement. In addition, the PRMC LED is designed so as to satisfy the cavity resonance condition, which maximizes SER, and for maximally constructive interference of the waves at the exit of the LED, which maximizes LEE. As a result, ignoring loss, the LEE and SER were found to be enhanced by factors of $LEE'/LEE=4.0$ and $SER'/SER=1.9$, respectively. Note that the SER results, computed by Eq.~\eqref{eq:SER_comp}, was benchmarked with literature results for the case of an emitter in vacuum~\cite{SER_FDFD}.

Figure~\ref{fig:power}(e) provides a better perspective on the PRMC dramatic PCE enhancement by plotting the radiation power in Figs.~\ref{fig:power}(a) and~(b) at one (free-space) wavelength above the structure. It is also interesting to observe in Fig.~\ref{fig:power}(b) that the radiation power pattern, both in the GaN and in the air, exhibits a series of discrete peaks, consistently with the above mentioned ray prediction that radiation constructive interference should occur at specific points on the LED surface. Finally, Fig.~\ref{fig:power}(f) compares the far-field radiation patterns of the two designs. The PRMC LED clearly produces a much higher directivity~(half-power beamwidth equals to $22.5^\circ$) than that the bare LED, which cannot be clearly defined given its absence of clear maximum. This is naturally due to the fact that the exit ray angles in the PRMC LED are normal to the structure whereas those in the bare LED are almost grazing on the LED surface. This feature of the PRMC LED may also be advantageous in LED applications requiring high directivity, such as for instance optical switches.

\begin{figure}
\centering\includegraphics[width=5cm]{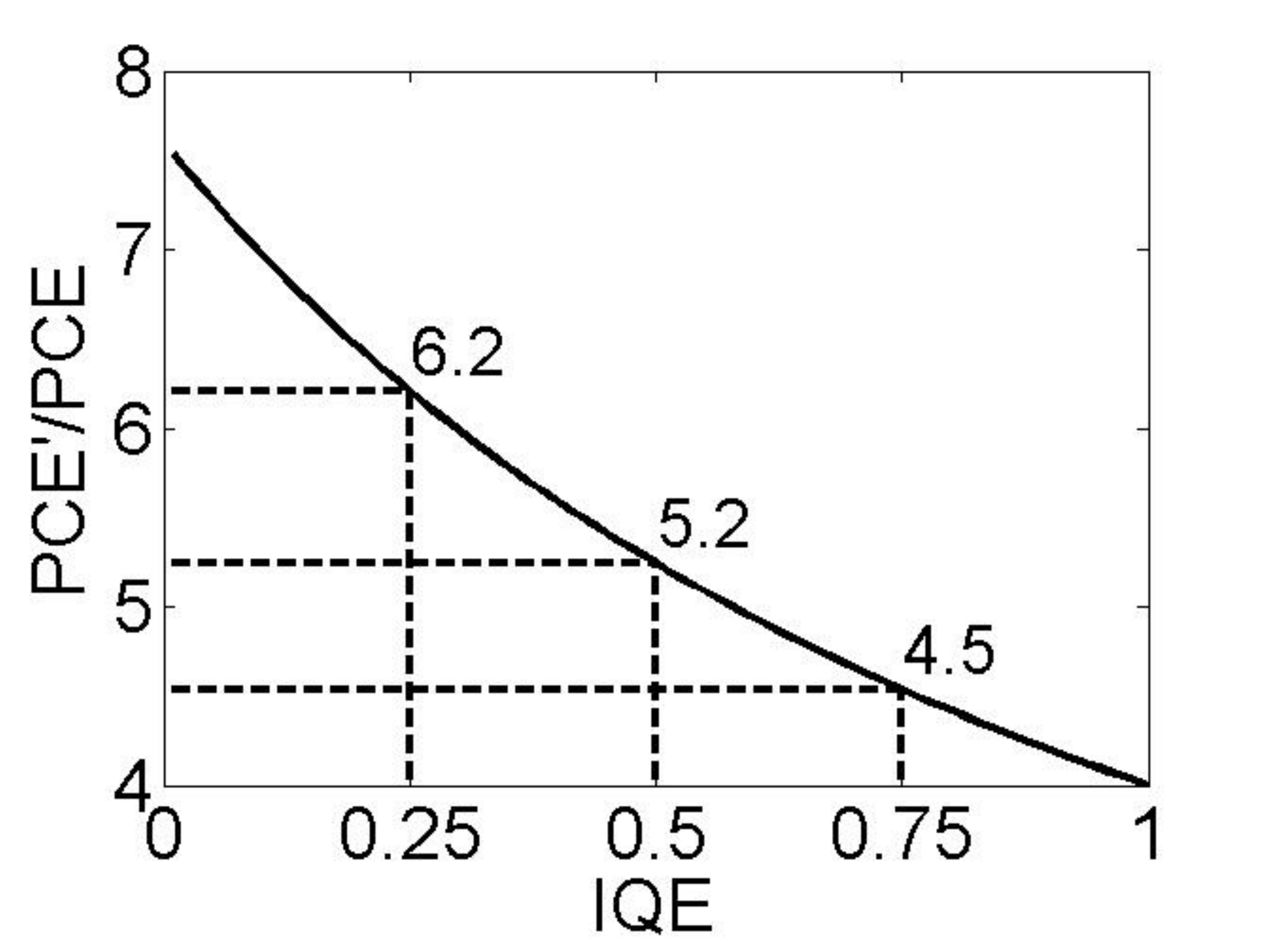}% Here is how to import EPS art
\caption{\label{fig:PCE_IQE}PCE enhancement factor of the PRMC LED with respect to the bare LED versus the IQE of the bare LED, computed by Eq.~\eqref{eq:PCE_enhancement} with $LEE'/LEE=4.0$ and $SER'/SER=1.9$.}
\end{figure}

Figure~\ref{fig:PCE_IQE} plots the PCE enhancement factor ($PCE'/PCE$) achieved in the PRMC LED design versus the IQE of the bare LED. It is seen that $PCE'/PCE$ ranges from 7.6 to 4.0 as IQE varies from 0 to 1, with values 6.2, 5.2 and 4,5 for IQEs of 0.25, 0.5 and 0.75, respectively. Naturally, as $IQE\rightarrow 1$, the PCE enhancement factor tend to be more and more due to $LEE'/LEE$, since the IQE enhancement possibility progressively reduces to zero in this limit.

The described design is a specific one, but it provides an idea on the PCE enhancement potential of the PRMC structure for LEDs. A practical LED has a light emitted layer with a size that is in the order of tens of wavelengths. Consequently, the modeling of a realistic LED would require a series of dipoles. In further studies, we will attempt to design a PRMC for such a series of dipoles, where the direct and indirect fields of the dipoles will be set as averaged effective fields.

\vspace{5mm}{\section{Conclusion}}

We have proposed and numerically demonstrated a novel light emitting diode (LED) architecture based on a partially-reflecting metasurface cavity (PRMC) structure. This PRMC structure is optimized, using a novel metasurface synthesis technique, to simultaneously maximize the light extraction efficiency~(LEE) and the spontaneous emission rate~(SER), and hence the power conversion efficiency~(PCE). Encouraging enhancement factors of 4.0 and 1.9 have been achieved for the LEE and SER, respectively, corresponding to PCE enhancement factors 6.2, 5.2 and 4.5 for IQEs of 0.25, 0.5 and 0.75, respectively, for a blue LED design. Beyond the interest in commercial LEDs, this paper presents the first double-metasurface cavity in the literature and, combining the physics of light extraction and spontaneous emission, it may additionally lead to deeper understanding of quantum-classical optics problems.

%%%%%%%%%%%%%%%%%%%%%%%%%%  body  %%%%%%%%%%%%%%%%%%%%%%%%%%

\vspace{5mm}\section*{Funding}
Collaborative Research and Development Project of the Natural Sciences and Engineering Research Council of Canada (NSERC) in partnership with the company Metamaterial Technologies Inc. (CRDPJ 478303-14).

\vspace{5mm}\section*{Acknowledgments}
Luzhou Chen acknowledges helpful discussions with Dr. Wei E. I. Sha and Mr. Yousef Vahabzadeh.

\end{document}